\documentclass[preprint,showpacs,preprintnumbers,amsmath,amssymb]{revtex4}

\usepackage{graphicx}
\usepackage{dcolumn}
\usepackage{bm}

\begin{document}


\title{The global build-up to intrinsic ELM bursts seen in
divertor full flux loops in JET}

\author{S. C. Chapman$^{1,3}$}
\email{S.C.Chapman@warwick.ac.uk}
\author{R. O. Dendy$^{2,1}$}

\author{T. N. Todd$^{2}$}
\author{N. W. Watkins$^{1,3,4,5}$}
\author{F. A. Calderon$^{1}$}

\author{J. Morris$^{2}$}

\author{JET Contributors}

\altaffiliation{See the Appendix of F. Romanelli et al., Proceedings of the 25th IAEA Fusion Energy Conference 2014, Saint Petersburg, Russia
}

\affiliation{EUROfusion Consortium, JET, Culham Science Centre, Abingdon, OX14 3DB, UK}

\affiliation{$^1$Centre for Fusion, Space and Astrophysics, Department of Physics, University of Warwick, Coventry, UK}%

\affiliation{$^2$ CCFE, Culham Science Centre, Abingdon, Oxfordshire OX14 3DB, UK}

\affiliation{$^3$Max Planck Institute for the Physics of Complex Systems, Dresden, Germany}
\affiliation{$^4$Centre for the Analysis of Time Series, London School of Economics, London, UK}

\affiliation{$^5$Dept. of Engineering and Innovation, Open University, Milton Keynes, UK}
\date{\today}

\begin{abstract}
 A global signature of the build-up to an intrinsic ELM is found in the phase of signals measured in full flux azimuthal loops in the divertor region of JET. Full flux loop signals provide a
 global measurement proportional to the voltage induced by changes
in poloidal magnetic flux; they are
 electromagnetically induced by the dynamics of spatially integrated current density. We perform direct time-domain analysis of the high time-resolution full flux loop signals  VLD2 and VLD3.
 We analyze plasmas where a steady H-mode is sustained over several seconds, during which all the observed ELMs are intrinsic;  there is
no deliberate intent to pace the ELMing process by external means. ELM occurrence times are determined from the Be II emission at the divertor. We previously \cite{chappop,chap4pager} found that
the occurrence times of intrinsic ELMs correlate with specific phases of the VLD2 and VLD3 signals.
 Here, we investigate how the VLD2 and VLD3 phases vary with time in advance of the ELM occurrence time. We identify a build-up to the ELM in which the VLD2 and VLD3 signals progressively align to the phase at which ELMs preferentially occur, on a $\sim 2-5 ms$ timescale.
 At the same time, the VLD2 and VLD3 signals become phase synchronized with each other, consistent with the emergence of coherent global dynamics in the integrated current density. In a plasma that remains close to a global magnetic equilibrium, this can reflect  bulk displacement or motion of the plasma.
This build-up signature to an intrinsic ELM can be extracted from a time interval of data that does not extend beyond the ELM occurrence time, so that these full flux loop signals could assist in ELM prediction or mitigation.
\end{abstract}
\pacs{52.27.Gr, 52.35.Mw,52.55.Fa}
\keywords{tokamak, synchronization, phase lock, edge localized mode}
\maketitle
\section{Introduction}
Enhanced confinement (H-mode) regimes in tokamak plasmas are characterized by  intense, short duration relaxation events known as edge localized modes (ELMs) \cite{intro1,intro2,intro3,intro4,intro5}.
Prevention of large amplitude ELMs is essential for ITER as each ELM releases particles and energy which load the plasma facing components; scaled up to ITER\cite{Haw2009}, the largest such loads would be unacceptable.  Theoretical \cite{peeling1,peeling2} and observational \cite{kstar} work suggests that the peeling-ballooning MHD instability of the plasma edge underlies ELM initiation, but as yet there is no comprehensive understanding of the sequence of physical processes involved in ELMing in terms of self-consistent nonlinear plasma physics.

Quantitative characterization of the dynamics of ELMing processes via their time domain properties, such as inter-ELM time intervals, or ELM event waiting times, is relatively novel \cite{deng,greenh,calderon,webster0,webster,murari} and has provided evidence of unexpected structure in the sequence of ELM occurrence times.
Recently \cite{chappop,chap4pager} we found that the signals from a system scale diagnostic, the full flux loops in the divertor region of JET, contain statistically significant information on the occurrence times of intrinsic ELMs: the ELMs tend to preferentially occur when the full flux loop signals are at a specific phase.
 Since the full flux loop signals capture aspects of the \emph{global} plasma dynamics
 including large scale plasma motion, this may suggest, as first proposed in \cite{chap4pager},
 a nonlinear feedback on a global scale where  the control system
and plasma behave as a single nonlinearly coupled system, rather than as driver and
response. This feedback may act to pace the intrinsic ELMs.

   In this paper we investigate the full time dynamics of the full flux loop signal phases, and directly test whether these signal phases contain  information on the build-up to an intrinsic ELM. We perform direct time domain analysis of  high time resolution signals from the  full flux loops in the divertor region in JET. These full flux loop VLD2 and VLD3 signals  are proportional to the
voltage induced by changes in poloidal magnetic flux. We use a simultaneous high time resolution Be II signal to determine the intrinsic ELM occurrence times.
We  focus on a sequence of JET plasmas that have a steady flat top for $\sim 5s$ and which all exhibit intrinsic ELMing in that there is no deliberate intent to control the ELMing process by external means. Importantly, the full flux loop signals have sufficiently large signal dynamic range, compared to the noise, to allow the time evolving instantaneous phase to be determined on timescales between one ELM and the next. ELMs tend to occur preferentially at a specific phase in the VLD2 and 3 signals. Here, we find that  the phases become progressively more strongly ordered from about $2-5 ms$ before the ELM up to the ELM time. Furthermore, the VLD2 and VLD3 signals become phase synchronized with each other during this build-up time. Global synchronized plasma dynamics is thus part of the build-up to an intrinsic ELM. The organization of the paper is as follows, in section 2 we introduce the data used in this study, in section 3 we describe how the full flux loop instantaneous phases are determined,  our main results are given in section 4, in section 5 we quantify the statistical significance of these results, and in section 6 we present a possible interpretation of these results following the scenario of \cite{chap4pager}. We provide significance tests  against null hypotheses, that is, phase alignment by chance coincidence, in the appendix.
\begin{figure}
\includegraphics[scale=0.55]{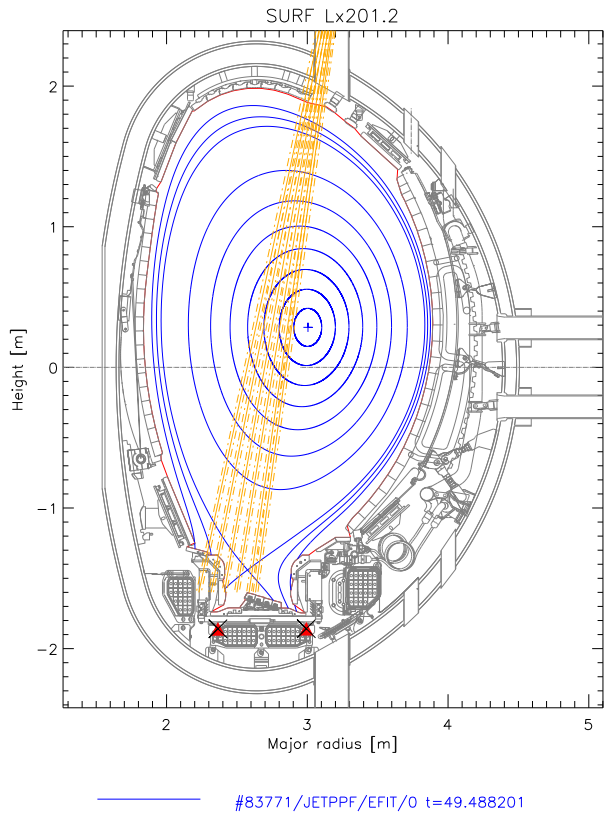}\vskip -2.5cm
\caption{Lines of sight for the BeII signal (yellow) and locations of the VLD2 and VLD3 full flux loops (red triangles) overplotted on EFIT
 magnetic surface reconstruction for JET plasma 83771 at t = 49.49 s.}
\label{surfpic}
\end{figure}

\begin{figure}
\includegraphics[scale=0.5]{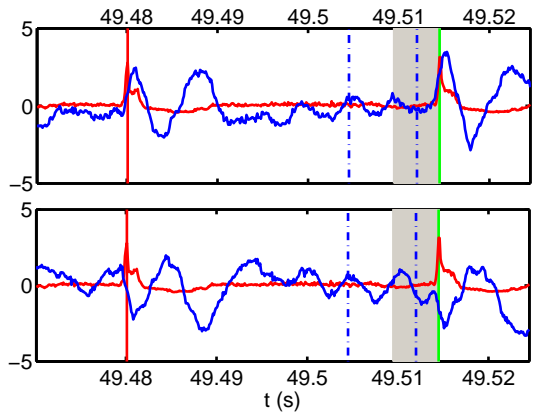}\vskip -2cm
\caption{
Standardized time traces for a pair of successive ELMs in JET plasma 83771.  Time traces of Be II intensity (red), with full flux loops (blue) VLD3 (upper panel) and VLD2 (lower panel).
 To facilitate comparison we have
 standardized the signal amplitudes by dividing by a multiple (10 for Be II, 2 for the VLD2 and 3) of their respective means over the flat-top H-mode duration, and then subtracted a local mean calculated over the interval denoted by the pair of vertical dot-dash blue lines.
The sign convention of the VLD2 and VLD3 signals is chosen such that they have opposite polarity.
The ELM occurrence times are indicated by vertical red and green lines.
 For reference the time interval between $0-5ms$ before the second ELM is shaded in grey.}\label{timeseries}
\end{figure}

\section{ELM and full flux loop time signatures}

We analysed the sequence of JET plasmas 83769-83775 discussed in \cite{chappop}. These are a subset of plasmas 83630-83794 analysed in \cite{webster}. Each has a flat-top H-mode duration of $\sim 5 s$. These all exhibit intrinsic ELMing in that there is no attempt to precipitate ELMs; the only externally applied time varying fields are those produced by the control system.
ELM occurrence times are inferred from
 the Be II signal, which we will compare  with measurements of
 the inductive voltage in the full flux loops VLD2 and  VLD3. These circle the JET tokamak toroidally at a location just below and outside the divertor coils. 
  The configuration of these diagnostics on JET is shown in Figure \ref{surfpic}.
The  signal voltage is induced by changes in poloidal magnetic flux through the surface encompassed by the loops.

\begin{figure}
\includegraphics[scale=0.55]{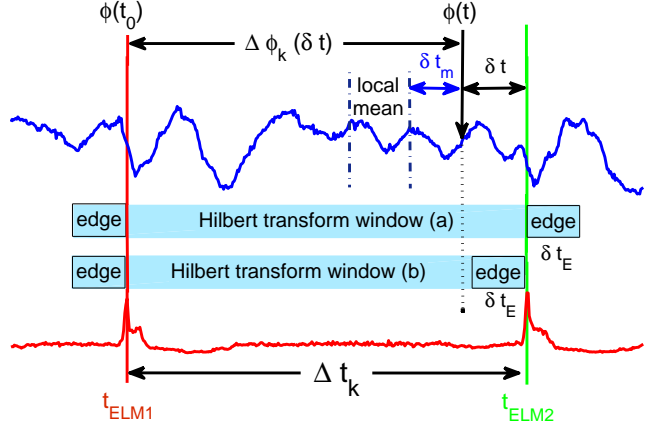}\vskip -2.5cm
\caption{Schematic (not to scale) of the procedure to determine the time dependent full flux loop phase difference $\Delta \phi_k(\delta t)$ as a function of the time interval $\delta t$ measured back from the time $t_{ELM2}$  of the second ELM in an ELM pair. $\Delta \phi_k(\delta t)$ is  w.r.t the phase at the time $t_{ELM1}$ of the first ELM. The phase can only be determined at times that are within the Hilbert transform window edge $\delta t_E$. One can then choose a Hilbert transform window (a) that goes beyond $t_{ELM2}$, so that $\delta t >0$ or (b) that stops at $t_{ELM2}$, so that $\delta t>\delta t_E$.
}\label{diagram}
\end{figure}

We  determine the ELM occurrence times $t_k$ by identifying the peak of the Be II signal within each ELM using the method described in \cite{chappop}.
From the occurrence times $t_k$ of these peaks,  the time intervals between successive ELMs, the ELM waiting times, $\Delta  t_k=t_{k}-t_{k-1}$ are found.
 In these plasmas there is time structure in the probability density of ELM waiting times. There is a lower cutoff in the ELM waiting time at $\Delta t \sim 10ms$, and there are time intervals where ELMs occur less often (\cite{chappop}, and in other plasmas, \cite{calderon}). Large ensemble statistical studies across many JET plasmas have also revealed \cite{webster,murari} that  the  ELM waiting time probability distribution shows time structure, that is, some ELM waiting times are more likely than others.

Signal traces for a representative pair of successive ELMs with a waiting time of $\sim 30 ms$ are shown in Figure \ref{timeseries} for plasma 83771.
 Following each ELM,
the figures show a characteristic large amplitude oscillatory  response in both of the full flux loop signals, the first cycle of which is on a timescale of $\sim  10ms$.
We previously identified \cite{chappop}  a class of prompt ELMs which are clustered approximately within  $10< \Delta  t <15ms$ and appear to be directly paced by this response to the previous ELM. These prompt ELMs will be excluded from the current analysis, here we consider ELMs that occur on longer timescales such that this initial  flux loop signal response to an ELM is seen to decay.
Intervals of quasi-periodic oscillations can be seen in the VLD2 and VLD3 signals throughout the time between one ELM and the next. We will now directly obtain the instantaneous phase of these signals in order to test for  information in these oscillations.

\begin{figure}\centering
\includegraphics[scale=0.6]{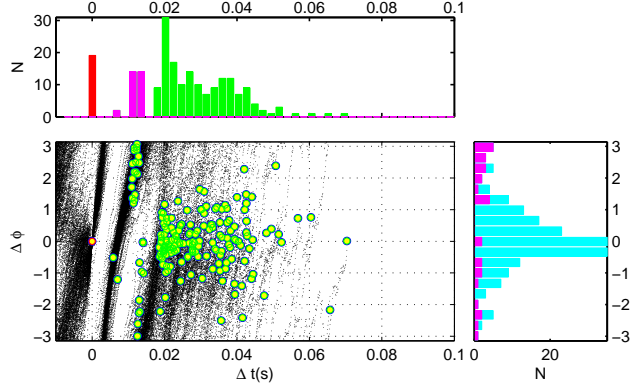}
\caption{ELM occurrence times and VLD2 phase shown for the  flat-top of JET plasma 83771.
The format of each set of panels is as follows:
Main panel: VLD2 instantaneous phase, modulo $2 \pi$, plotted as a function of time following each ELM up to the occurrence time of the next ELM.
 The  coordinates  are time $\Delta t=t-t_{0}$ and phase difference $\Delta
  \phi=\phi(t)-\phi(t_0)$, where $t_0=t_{ELM1}$. ELM occurrence times are marked on each VLD2 trace with yellow filled red circles (first ELM) and green circles (second ELM).  Right hand panel: histogram of VLD2 $\Delta \phi$ at the time of all the second ELMs (blue), overplotted (pink) for the prompt ELMs with  waiting times $\Delta  t <15ms$.
Top Panel: histogram of ELM occurrence times $\Delta t=t-t_{0}$ for the first ELM (red) and
 the second ELMs (green), overplotted (pink) for the prompt ELMs. The frequency $N$ of first ELM times has been rescaled by $1/10$.
}\label{phasetime}
\end{figure}

\section{Determination of full flux loop instantaneous phase}

A time series $S(t)$ has a corresponding
analytic signal defined by $S(t)+iH(t)=Aexp[i\phi(t)]$, where $H(t)$ is the Hilbert transform of $S(t)$, defined in \cite{gabor,text1,pikbook} see also \cite{phaseprl1,phaseprl2}.
This defines an instantaneous amplitude $A(t)$  and phase $\phi(t)=\omega(t) t$ where  the instantaneous  frequency is $\omega(t)$ for the real signal $S(t)$.
We compute the analytic signal by Hilbert transform over each waiting time $\Delta t_k$ between each pair of ELMs. The procedure is summarized in the schematic shown in Figure \ref{diagram}, which shows the domain over which the Hilbert transform is calculated relative to a pair of ELMs occurring at $t_{k-1}=t_{ELM1}$ and $t_{k}=t_{ELM2}$. We will obtain the phase for the full flux loop signals for a sequence of times $\delta t$ preceding the second ELM of each pair, that is, at times  $t_{ELM2}-\delta t$. We will need to choose a zero time $t_0$ to define a phase difference in the full flux loop signals $\Delta \phi=\phi(t)-\phi(t_0)$; here $t_0$ will be an estimate of the occurrence time of the first ELM.

The full flux loop signals are sufficiently above the noise that we can use this method to determine their instantaneous phase. The instantaneous phase cannot be directly extracted for the Be II signal because its noise level is usually too high.
We first perform a  3 point spline smoothing on the VLD2 and VLD3  time series to remove noise fluctuations on the sampling timescale. The signal analyzed must oscillate about zero in order for the instantaneous phase to be well determined from the analytic signal, we can ensure this locally by subtracting a locally determined mean specified as shown in Figure \ref{timeseries}. The signal local mean is determined within a window that is shifted back in time by $\delta t$, in the results shown here we used a window
$T_A=[t_{k}-0.01, t_{k}-0.0025]s-\delta t$,
 relative to each ELM occurrence time $t_k$ (so that on the schematic $\delta t_m=2.5 ms$).
The Hilbert transform has a single-sided Fourier transform which is approximated via fast Fourier transform over the finite time window of the data. We therefore need to use a time window that is larger than that of the time domain of interest to avoid edge effects, so that we only calculate the instantaneous phase at times within a window edge time interval $\delta t_E$ of the ends of the time window of data. We have found that in these time-series a $\delta t_E >1 ms$ is sufficient to give stable values of the instantaneous phase and all results presented here use this value of $\delta t_E$. We have varied $T_A$ $\delta t_m$, and $\delta t_E$ to check the robustness of our results.

The above methods are only effective if the full flux loop signals have good signal/noise, do not have too large a dynamic range in  response to all the ELMs, and if the mean of the signal does not vary too rapidly. The high rate of change of instantaneous phase with time of the full flux loop signals  requires well defined ELM occurrence times in order to cleanly determine any phase relationship.

\begin{figure}\centering
\includegraphics[scale=0.7]{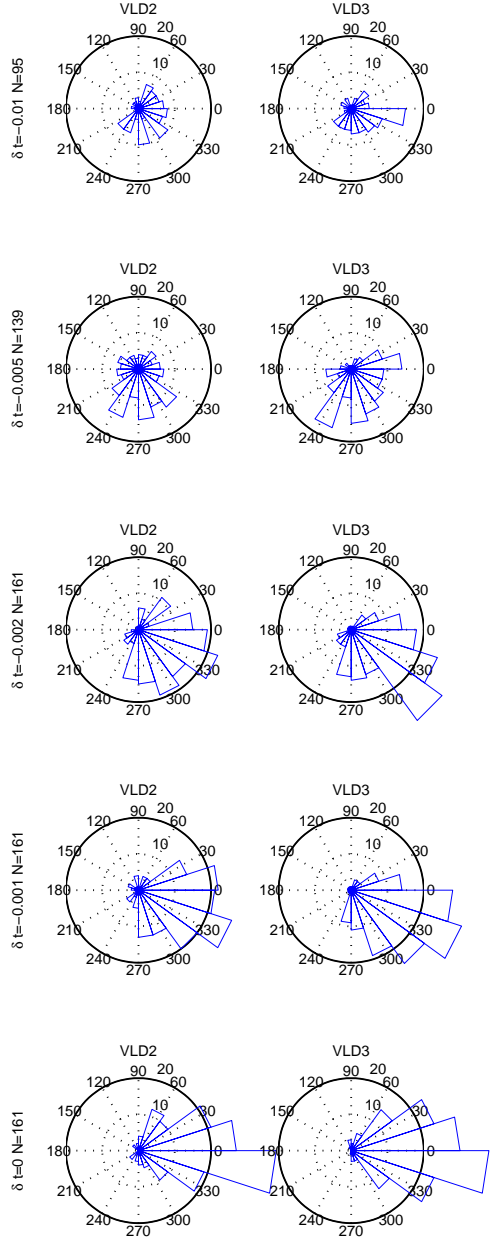}
\caption{Polar plots of histograms of VLD2 (left panels) and VLD3 (right panels) phase difference $\Delta \phi$ just before an ELM in the flat-top of JET plasma 83771. The phase difference  is calculated from the time of the first ELM to a time $\delta t$ before the second ELM so that $\Delta \phi=\phi(t_{ELM2}-\delta t)-\phi(t_{ELM1})$. From  bottom to top $\delta t=[0$ $1$ $ 2 $ $5 $ $10]ms$.
The histograms include all ELM pairs with  waiting times $\Delta  t >15ms-\delta t$, the number $N$ in the histogram decreases with increasing $\delta t$.
Hilbert transform time window (a) is used to determine the VLD phases and it extends $\delta t_E=1ms$ beyond $t_{ELM2}$, the time of the second ELM. The interval used to determine the VLD signal means just before the ELM moves back with $\delta t$.
}\label{clockall}
\end{figure}

\begin{figure}[h]\centering
\includegraphics[scale=0.7]{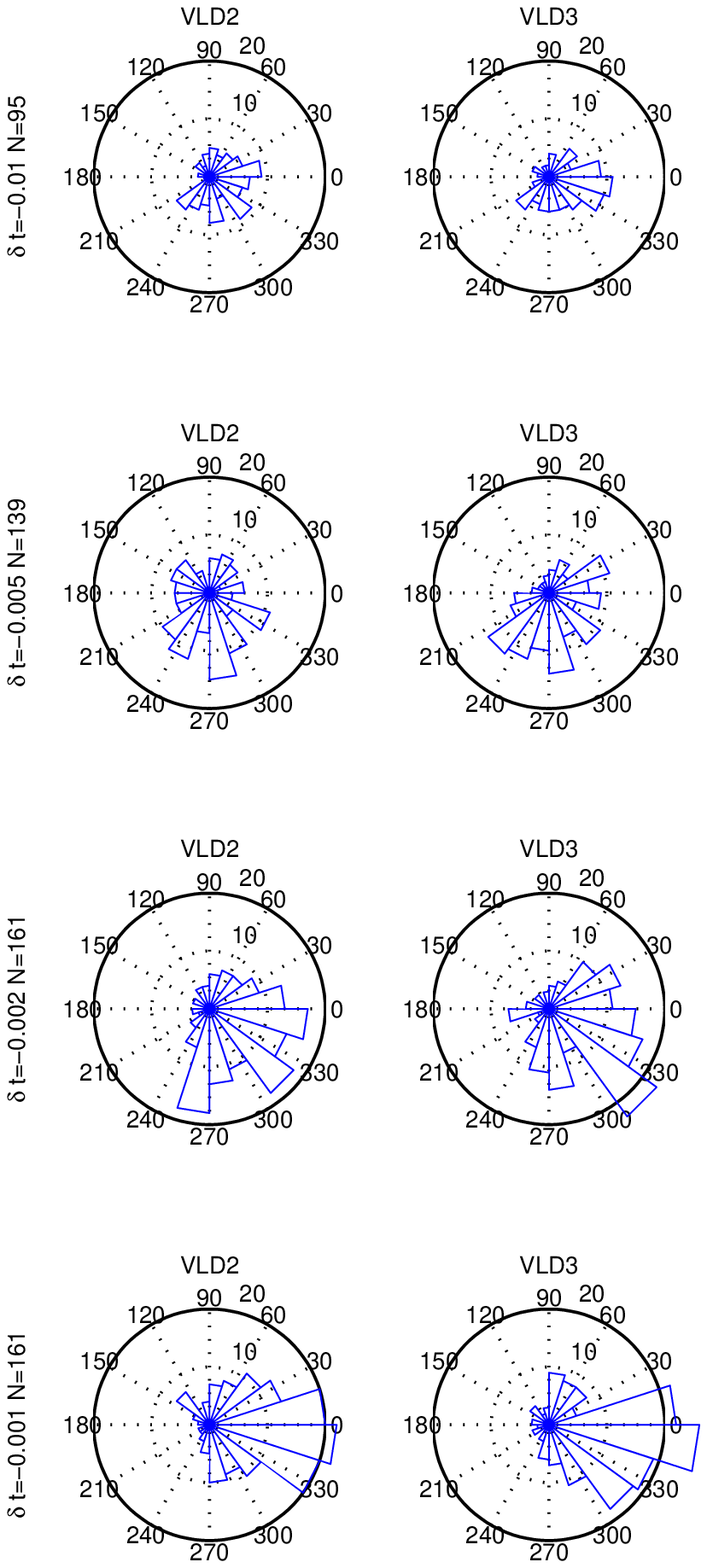}
\caption{Polar plots of histograms of VLD2 (left panels) and VLD3 (right panels) phase difference $\Delta \phi$ just before an ELM in the flat-top of JET plasma 83771. The same data and analysis is used as in Figure \ref{clockall} except that now Hilbert transform time window (b) is used to determine the VLD phases and it stops at $t_{ELM2}$, the time of the second ELM. Phase differences can then only be determined for times $\delta t>\delta t_E=1ms$ before $t_{ELM2}$.  From  bottom to top $\delta t=[1 $ $2 $ $5 $ $10]ms$.
}\label{clock1back}
\end{figure}
\section{Full flux loop instantaneous phase and build-up to an ELM}

We will first discuss results using Hilbert transform window (a) which extends beyond the time $t_{ELM2}$ of the second ELM, so that we can obtain the instantaneous phase at times up to the ELM occurrence time.
In Figure \ref{phasetime} we plot the instantaneous phase of the full flux loop signal versus time for all the ELMs in JET plasma 83771.  The main figure panel plots time from $t_0$, that is, $\Delta t=t-t_0$ versus the instantaneous phase difference $\Delta \phi=\phi(t)-\phi(t_0)$ of the VLD2 signal. In Figure \ref{phasetime} we set $t_0=t_{ELM1}$, the time of the first ELM as determined from the Be II signal.
The first (red circle) and second (green circle) ELM times, as determined from the Be II signal, are  overplotted on each corresponding VLD2 trace. On the plot, the first ELM  has coordinates $\Delta t=0$ and $\Delta \phi=0$ by definition. The coordinates of the second ELM  are the waiting time
$\Delta t_k=t_{ELM2}-t_{ELM1}$ and corresponding phase difference $\Delta \phi_k=\phi(t_{ELM2})-\phi(t_{ELM1})$.
 Histograms are shown of the waiting times $\Delta t_k$ (top panel) and phase differences $\Delta \phi_k$ (right panel) for all the $k=1..N$ ELM pairs. There is a group of prompt ELMs \cite{chappop} with $\Delta t<15ms$, indicated by pink bars, which are distinct in both arrival time and phase (as they are a response to a transient, their phases are not well determined by the Hilbert transform method). We have previously identified these prompt ELM events as being directly correlated with the response to the previous ELM and will exclude them from the  analysis to follow by only considering ELM pairs with waiting times $\Delta t>15ms$.
  These ELMs, with $\Delta t>15ms$, are phase bunched with a peak around zero phase. We obtain the same results for the VLD3 and for the other plasmas in the sequence.

 ELMs are thus more likely to occur when the full flux loop signals are at a specific phase w.r.t. that of the preceding ELM. Prompt ELMs occur within the coherent (in amplitude and phase) response to the previous ELM which can clearly be seen in the full flux loop signals \cite{chappop}. For all other, non-prompt ELMs, the full flux loop signals do not remain coherent in both amplitude and phase throughout the inter-ELM time interval. The question is then whether there is detectable phase coherence at all times (implying that the system retains a memory of the preceding ELM) or whether phase coherence is lost, and then re-emerges as part of the build-up to the next ELM.
 Figure \ref{clockall} shows polar plots of the histogram of the phase differences $\Delta \phi_k (\delta t)$ for all of the ELM pairs in plasma 83771, The phase difference is determined at time  $t=t_{ELM2}-\delta t$, that is, at time $\delta t$ before the second ELM. As the preceding ELM generates a large amplitude response in the full flux loop signals on a timescale $\sim 10ms$ we will exclude ELM pairs with waiting times  $\Delta  t <15ms+\delta t$; hence the number $N$ of samples in the histogram decreases with increasing $\delta t$.
  As in Figure \ref{phasetime} we use Hilbert transform window (a) which extends beyond the time $t_{ELM2}$ of the second ELM, so that we can obtain the instantaneous phase at times up to the ELM occurrence time.  The bottom panels in Figure \ref{clockall} are at the time of the ELM, $\delta t=0$ so that the bottom left hand plot is a polar histogram of the same data as in the right hand panel of Figure \ref{phasetime}. The time before the ELM $\delta t$ increases moving up the plot. We then see that the phase difference qualitatively becomes progressively more ordered from $\delta t \sim 5ms$ and that there is a clear phase bunching after $\delta t \sim 2ms$. This suggests that there is a signature of the build-up to an ELM in the full flux loop signals and we will quantify the degree of phase bunching in the next section.

Figures \ref{phasetime} and \ref{clockall} required a Hilbert transform window (a) which extended beyond the time $t_{ELM2}$ of the second ELM so that the instantaneous phase at times up to the ELM occurrence time could be calculated. As a check on the robustness of the ELM build-up signature we repeat the analysis with Hilbert transform window (b) which stops at the time $t_{ELM2}$ of the second ELM so that only information before the ELM occurrence time is used. The resulting polar histograms are shown in Figure \ref{clock1back}, where apart from the different Hilbert transform window, the data and analysis is the same as that used to produce Figure \ref{clockall}. Now, we can only consider times before the second ELM $\delta t>\delta t_E=1ms$. Nevertheless we still see in these histograms a clear phase bunching on the same timescales  as in Figure \ref{clockall}, where information from times beyond the ELM time  $t_{ELM2}$ was used.

\begin{figure}\centering
\includegraphics[scale=0.7]{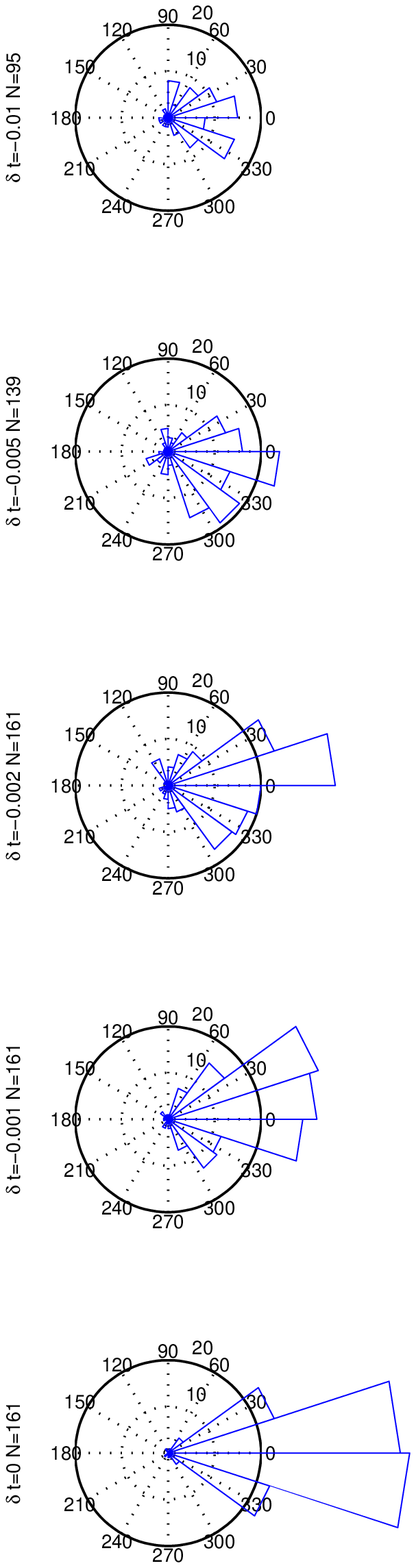}
\caption{Polar plots of histograms of the instantaneous phase difference between the VLD2  and VLD3 signals just before an ELM in the flat-top of JET plasma 83771. The phase difference  is calculated directly between the two signals,
so that $\Delta \phi=\phi(VLD2(t))-\phi(VLD3(t))$. The format is the same as in Figure \ref{clockall}. Hilbert transform window (a) is used to determine the phases.
}\label{clockdiff}
\end{figure}

The above results test for temporal coherence in the build-up to an ELM, that is, over what time interval do we always see the same phase in the VLD2 or VLD3 just before an ELM.
The VLD2 and VLD3 full flux loops are both in the divertor region of JET and in Figure \ref{timeseries} we can see that they are very similar in their time variation, however they are not identical. The time evolving phase difference between these signals provides a measure of spatial coherence, that is, coherent large scale plasma motion in the region of these full flux loops will tend to make their phases align. We test this idea in Figure \ref{clockdiff} where we plot polar histograms of
 the instantaneous phase
difference between the VLD2 and VLD3 signals at times $\delta t$ before each of the ELMs in plasma 83771, in the same format as Figure \ref{clockall}.
From the top panel we see that their phase difference at all times shows some alignment, it is within $\sim \pm 60 $ degrees of its mean at $\delta t=10ms$ before the ELM. However again for times $\delta t<5ms$, that is, just before the ELM, we see that the phase difference in these two signals tends to zero, that is, they become phase synchronized.

\section{Circular statistics and the Rayleigh test}

We use the Rayleigh test and associated circular statistics (see e.g. \cite{fisher,berens} and refs. therein) to quantify the extent to which the phase differences are aligned, and the statistical significance of any such alignment. Using the procedure described above, we determine the phase differences $\Delta \phi_k$ for the $k=1..N$ ELM pairs in a given plasma.
If each phase is represented by a unit vector  $ \underline{r}_k=(x_k, y_k)=(cos \Delta \phi_k, sin \Delta \phi_k)$ then a measure of their alignment is given by the magnitude of the vector sum, normalized to $N$. This is most easily realized if we use unit magnitude complex variables to represent the $ \underline{r}_k=e^{i\Delta \phi_k}$. Then if
\begin{equation}
\sum_{k=1}^N \underline{r}_k=re^{i \bar{\phi}}
\end{equation}
the Rayleigh number is the magnitude of the sum:
\begin{equation}
R=\frac{1}{N}\mid\sum_{k=1}^N \underline{r}_k\mid=\frac{r}{N}
\end{equation}
and the mean phase angle is $\bar{\phi}$.
Clearly, if $R=1$ the phases are completely aligned, however $R=0$ does not distinguish random alignment from ordered anti-alignment. We will consider two other statistics here. The first is an estimate of how closely aligned the phases are with the mean phase angle. We can calculate centred trigonometric  moments relative to the mean phase angle $\bar{\phi}$:
\begin{equation}
m_q=\frac{1}{N}\sum_{k=1}^N e^{i q (\Delta \phi_k-\bar{\phi})}=r_q e^{i  \delta \phi_q}
\end{equation}
 We will consider $q=2$, then the phase angle of $m_2^{\frac{1}{2}}$, that is $\delta \phi_2 /2$ is a measure of the angular variance around the mean $\bar{\phi}$;  this can take values $[0-\pm\pi]$. We will plot  this quantity as a standardized, positive definite,  angular variance $\sigma_\phi=\mid \delta \phi_2\mid/2\pi$ so that $\sigma_\phi$ is in the range $[0,1]$.

\begin{figure}\centering
\includegraphics[scale=0.4]{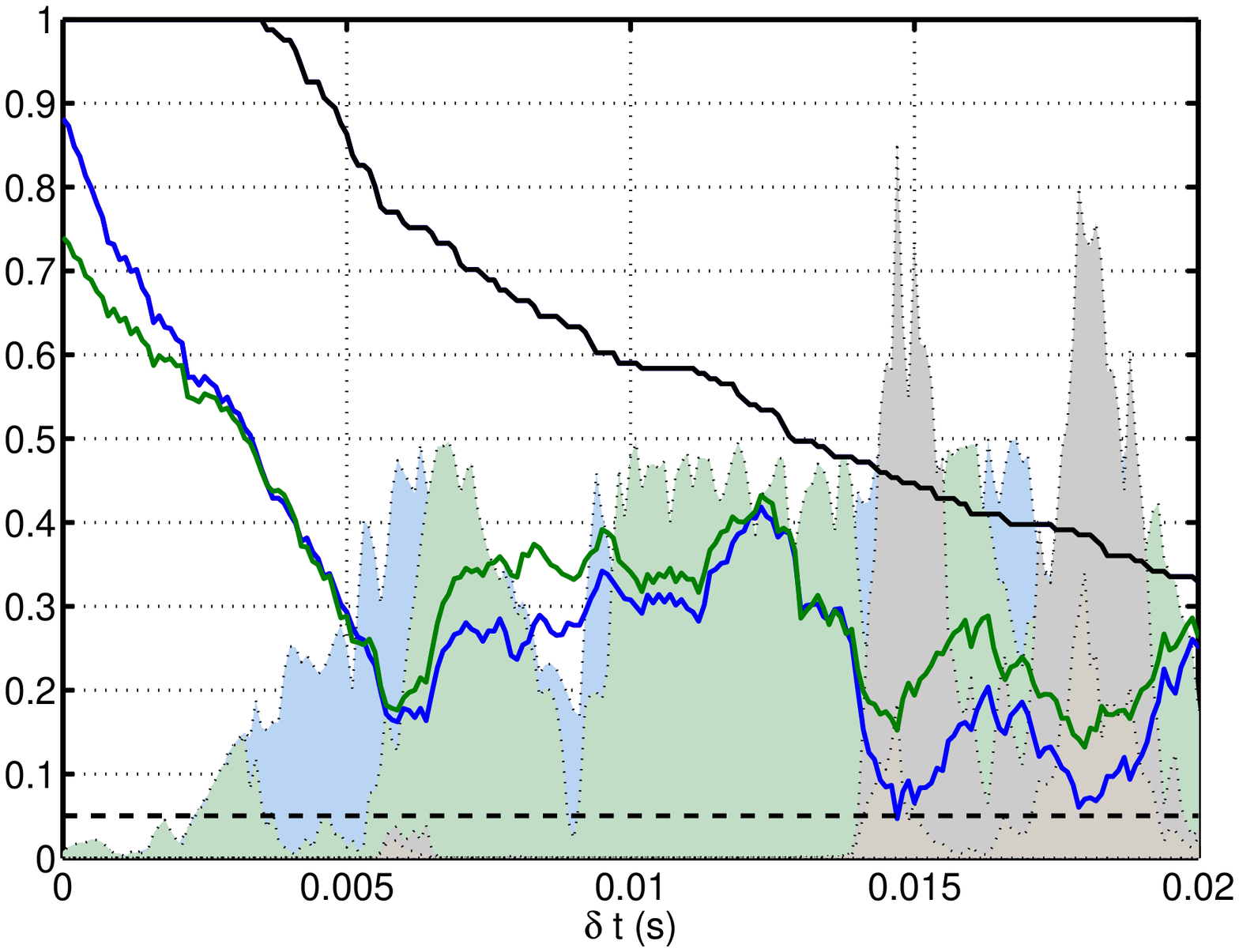}
\caption{Rayleigh statistics for VLD2 phase difference $\Delta \phi$ just before an ELM in the flat-top of JET plasma 83771.
The phase difference  is calculated from  time $t_0$ to a time $\delta t$ before the second ELM
so that $\Delta \phi=\phi(t_{ELM2}-\delta t)-\phi(t_{0})$. The figure plots variation with $\delta t$ ($x$ axis) of Rayleigh $R$ for $t_0=t_{ELM1}$ (green) and for $t_0=t_{VLDmin}$
(blue). The corresponding standardized angular variance is plotted as light blue and green shading.
The corresponding \emph{p}-values are indicated by the dark and light grey shading respectively. The $p=0.05$ level is indicated by the horizontal dashed black line.
The sample includes all ELM pairs with  inter-ELM time intervals $\Delta  t >15ms+\delta t$, the number $N$ in the sample decreases with increasing $\delta t$; the fraction $N(\delta t) / N(\delta t=0)$ of ELM pairs in the sample is plotted (black). Hilbert transform window (a) is used.
}\label{elmallR}
\end{figure}
\begin{figure}\centering
\includegraphics[scale=0.4]{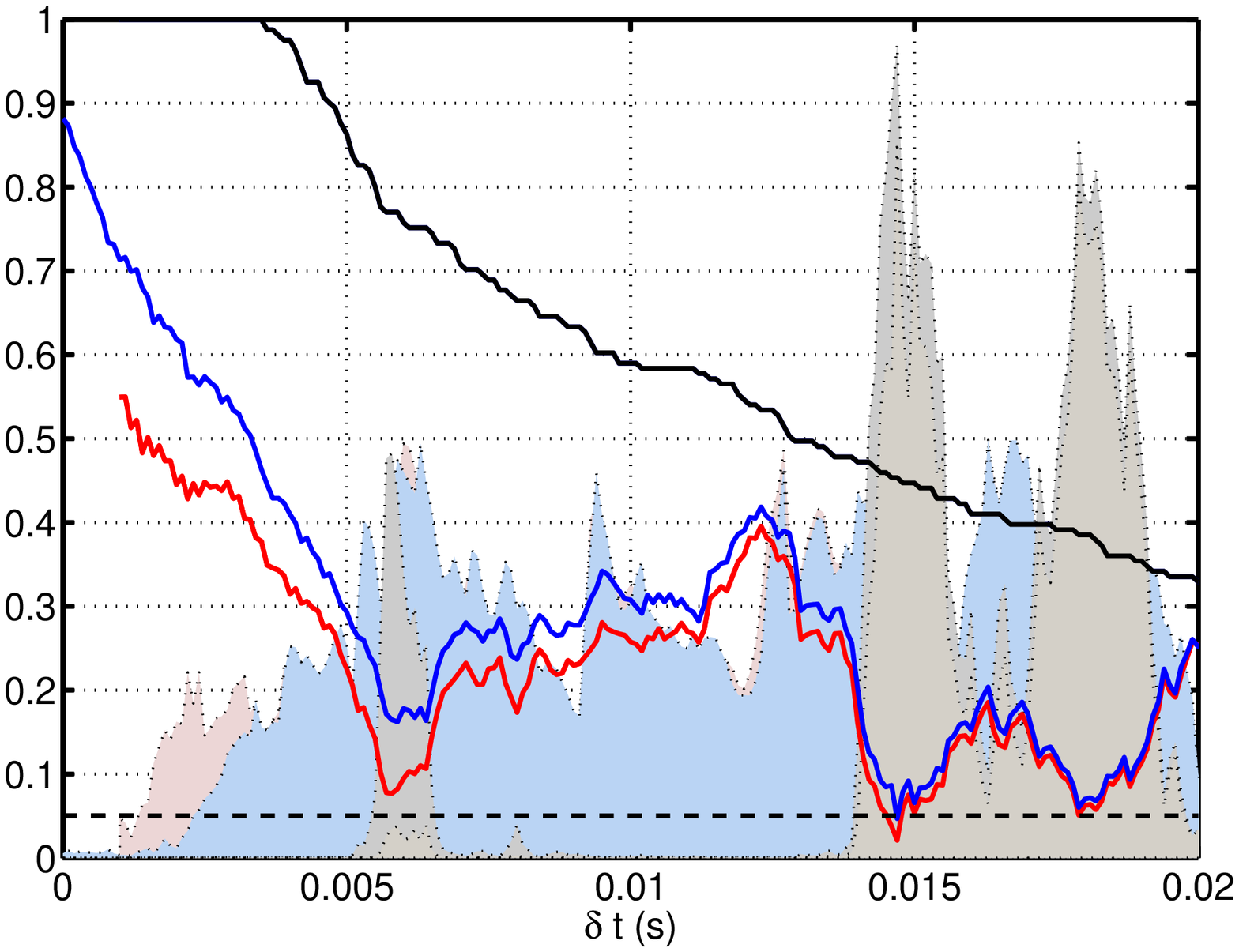}
\caption{Rayleigh test for VLD2 phase difference $\Delta \phi$ just before an ELM in the flat-top of JET plasma 83771. The format is as in the previous figure.
The phase difference  is calculated from  time $t_0$ to a time $\delta t$ before the second ELM
so that $\Delta \phi=\phi(t_{ELM2}-\delta t)-\phi(t_{0})$ where $t_0=t_{VLDmin}$. The blue line is the same as the previous figure, Hilbert transform window (a) is used to determine the phases. The red line is the same analysis using Hilbert transform window (b). These time windows end at
$t_{ELM2}+1ms]$ (blue) and $t_{ELM2}]$ (red) respectively. Normalized angular variance shown in shaded light blue and red respectively.
The corresponding \emph{p}-values are indicated by the  light and dark grey shading respectively.
}\label{elm1backR}
\end{figure}
 The second is an estimate of the \emph{p}-value under the null hypothesis that the vectors are uniformly distributed around the circle which is given by:
\begin{equation}
p=exp\left[\sqrt{1+4N+4N^2(1-R^2)}-(1+2N)  \right]
\end{equation}
so that a small value of \emph{p} indicates significant departure from uniformity, i.e. the null hypothesis can be rejected with 95\% confidence for $p<0.05$.

We now calculate the Rayleigh $R$, the standardized angular variance,  and $p$ values as a function of the time before the ELM $\delta t$,
 corresponding to the polar histograms above. Figure \ref{elmallR} corresponds to the analysis of Figure \ref{clockall}, where we have used Hilbert time window (a)to obtain the
 VLD2 phase difference $\Delta \phi_k$ at times $\delta t$ just before each ELM.  The phase difference  is again  calculated from the time of the first ELM to a time $\delta t$ before the second ELM so that $\Delta \phi_k=\phi(t_{ELM2}-\delta t)-\phi({t_0})$. On Figure \ref{elmallR} the
 green line is the Rayleigh $R$ for the analysis of Figure \ref{clockall} where we calculate the phase differences from the zero time at the first ELM $t_0=t_{ELM1}$.  The times of the extrema of the characteristic initial large amplitude oscillatory  response to an ELM, which is seen in both the full flux loop signals, have been found \cite{chappop} to  provide a better determined zero time $t_0$. The blue line in Figure \ref{elmallR} is the $R$ obtained for $t_0=t_{VLDmin}$, the first minimum of the VLD2 signal following the preceding ELM. We can then see that $R>0.3$ for $\delta t<5ms$ before the ELM occurs, and systematically increases as we approach the ELM occurrence time. Within this time interval  the standardized angular variance $\sigma_\phi$ is small, it gradually increases with $\delta t$ as the phases become more disordered. The \emph{p-} statistic remains small for times $\delta t<15ms$ indicating that the
distribution of phases remains far from circular. However this is not a smooth trend, there are short intervals (for example around $\delta t\sim 6ms$ where $p\sim 0.05$. We have found that for $\delta t>5ms$ the details of where short-lived fluctuations in $R$, $\sigma_\phi$ and \emph{p-}value occur are not robust, they vary with the dataset and with the detailed parameters of how the Hilbert transform is computed. However the overall trends are robust, in particular, alignment of the phases around a single value for $\delta t<5ms$, that is, large $R$, and small $\sigma_\phi$ and \emph{p-}value.

In Figure \ref{clock1back} we only used signals up to, and not beyond, the time of the ELM in order to test for the ELM build-up signature in the full flux loop phases. The corresponding circular statistics are plotted in Figure \ref{elm1backR} where the phase differences are obtained for $t_0=t_{VLDmin}$.
The blue line in Figure \ref{elm1backR} replots that in Figure \ref{elmallR}, it is calculated using Hilbert transform time window (a) which extends to times beyond the ELM occurrence time. The red line in Figure \ref{elm1backR} is obtained using the same analysis and data, but with phases calculated using  Hilbert transform time window (b) which stops at the ELM occurrence time. We can see that the build-up to an ELM in the full flux loops can still be resolved only using information from before the ELM occurrence time.

\begin{figure}\centering
\includegraphics[scale=0.4]{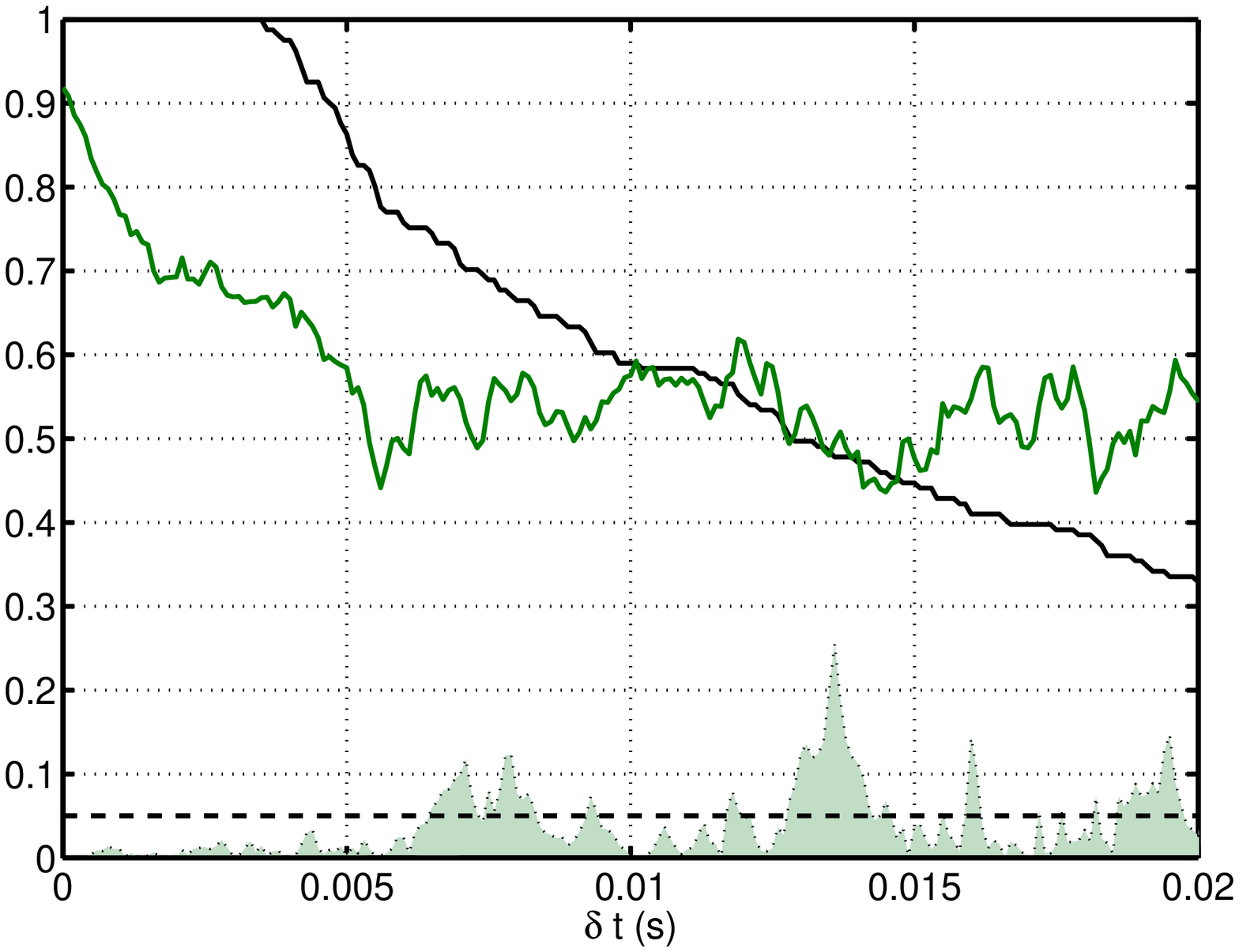}
\caption{Rayleigh test for the instantaneous phase difference between the VLD2 and VLD3 signals just before an ELM in the flat-top of JET plasma 83771. The format is as in the previous figure.
The phase difference  is calculated directly between the two signals,
so that $\Delta \phi=\phi(VLD2(t))-\phi(VLD3(t))$. The green line is the $R$ value, Hilbert transform window (a) is used to determine the phases.  Normalized angular variance shown in shaded light green and the \emph{p}-values  by the  light shading, here it is too small to be visible on the plot.
}\label{clockdiffR}
\end{figure}

Finally, in Figure \ref{clockdiffR} we plot the Rayleigh statistics for the instantaneous phase
difference between the VLD2 and VLD3 signals that was shown in
 Figure \ref{clockdiff}. These signals are very similar in their time variation as can be seen in Figure \ref{timeseries},
however they are not identical. From Figure \ref{clockdiffR} top panel we see that their phase difference at all times shows some alignment, it is within $\sim \pm 60 $ degrees of its mean so that $R\sim 0.5$ in Figure \ref{clockdiffR}. However again for times $\delta t<5ms$, that is, just before the ELM, we see that the phase difference in these two signals tends to zero, that is, they become phase synchronized.

We have quantified the values that these circular statistics can take for these time-series due to chance coincidence. Chance coincidence can occur between time-series that have non-trivial time-structure, for example,  roughly periodic ELM occurrence times may preferentially occur at specific phases of a roughly sinusoidal signal. We have
 constructed a set of surrogate time-series and repeated the above analysis to explore this possibility. This is described in detail in the appendix, and establishes that the phase alignments seen for times $\delta t<5ms$ can be distinguished as statistically distinct from chance occurrence and thus are evidence for correlation.

  Our main result is  that there is a  signature of the build-up a non-prompt ELM in the phases of the full flux loop signals. ELMs tend to occur preferentially at a specific phase in the VLD2 and 3 signals. The $R$, $\sigma_\phi$ and \emph{p-}values are all consistent with alignment of the phases from about $2-5 ms$ just before the ELM occurs. Furthermore, the VLD2 and VLD3 signals become phase synchronized with each other during this build-up time. Global, spatio-temporally  synchronized plasma dynamics is thus part of the build-up to an intrinsic ELM. We cannot detect statistically significant
phase coherence at all times whereas we do detect phase coherence re-emerging as part of the build-up to the next ELM.

\section{Discussion}

Our results rely upon a new approach to the analysis of an existing JET diagnostic, the full flux loops  VLD2 and VLD3 signals, alongside ELM timings from the Be II signal.
We have performed direct time-domain analysis of  high time resolution full flux loop signals arising from  the dynamics of spatially integrated current density,
 with high time resolution determination of the ELM timings. In addition to the main results of this paper, we will in this section develop our recent conjecture \cite{chap4pager} in the light of these results. Our aim is to frame a testable hypothesis for future work.

It is well established experimentally that ELMs can be triggered by applied magnetic 'kicks' delivered by the vertical stabilization control coils which drive vertical plasma movement, including global changes in the divertor region.
    In such triggering experiments in JET,  ELMs preferentially occur when the plasma is in a specific phase in its vertical motion (downwards), and delays of $\sim 2-3ms$ typically are observed between the start of the kick and the ELM\cite{sarkick,Luna2009}. Similar behaviour, i.e. ELM occurrence when the plasma is in a specific phase in its vertical motion, is seen in other devices, e.g. \cite{langppcf,nstx} and references therein.  Furthermore the  velocity perturbation associated with intrinsic ELMs is found to set a minimum threshold value that must be exceeded in order to trigger ELMs with the vertical coils \cite{langppcf}. The build-up phase to an ELM that has been magnetically kicked thus involves global plasma motion at a specific phase. This global plasma displacement can then modify conditions at the plasma edge, such that peeling-ballooning and perhaps other instabilities become active, leading to the ELM burst. The details are complex and may be device dependent \cite{kim09}; but  the essential point here is that the kicked ELM burst follows a global perturbation in the plasma dynamics and occurs at a specific phase thereof.

We have presented evidence for the emergence of coherent global dynamics in the integrated current density in the $\sim 2-5 ms$ build-up to an \emph{intrinsic} ELM. In a plasma that remains close to a global magnetic equilibrium, this can reflect  bulk displacement or motion of the plasma.
We see this build-up in the full flux loop signals which track the dynamics of the integrated current density in the divertor region. The VLD2 and VLD3 signals become phase synchronized during this build-up, suggesting a spatially coherent large-scale plasma perturbation. The intrinsic ELMs  are found to preferentially occur at a specific phase in the full flux loop signals, that is, at a specific phase in this global perturbation in the plasma. If this global perturbation is sufficient to modify conditions at the plasma edge to favour instability, then an intrinsic ELM can occur.

Our results  suggest one possible  scenario for intrinsic ELMing where   the plasma and its interacting environment together self-generate a global plasma perturbation, such that  the plasma is magnetically 'self-kicked', which then leads to an ELM.
Self-generation of global motion could occur via
 nonlinear feedback between  the multiscale dynamics of the plasma and its interacting environment, including the control system, as we first suggested in \cite{chap4pager}. The steady state of the JET flat top
plasmas  is actively maintained by perturbations from the control system reacting
to plasma motion. Integrated over the largest spatial scales, the reaction of the plasma to these
perturbations is seen in the full flux loop signals. These signals reflect the control system
and plasma behaving as a single nonlinearly coupled system, rather than as driver and
response.
If there were coupling between the global plasma environment, including the control system, and each of several growing modes in the plasma, these modes could become  synchronized\cite{pikbook,phaseprl1,phaseprl2}, through their individual interactions with the global plasma/control system environment,  without the need of coupling between the modes themselves. Large scale plasma motion would then develop on  timescales characteristic of the  dynamics  of the global plasma environment. We have found an ELM build-up timescale of $\sim 2-5ms$, which is similar both to the $\sim 2ms$ time constant of the known unstable mode in
 the vertical control system on JET \cite{Neto2011}, and the $\sim 2-3ms$ response time to generate global plasma motion from active kicks in the vertical stabilization control coils \cite{sarkick,Luna2009}.

 The VLD2 and VLD3 full flux loops also capture the initial integrated plasma and control system response to an ELM \cite{chappop}.
 If this integrated plasma and control system response again corresponds to global plasma motion, it may be expected to act as a 'kick' to directly trigger an ELM, if this global perturbation is sufficient to modify conditions at the plasma edge for instability.  We
   found  \cite{chappop} that \emph{prompt} ELMs sometimes occur at a specific phase within this initial response to the previous ELM. This suggests an additional testable hypothesis: that compound ELMs are a pattern of successive  prompt ELMs and again arise from global plasma motion emerging as above. This is consistent with the observation \cite{calderon} of a narrow spread in the time intervals between successive component ELMs in a compound ELM sequence. We would then expect to see a well-defined phase relationship between high time resolution full flux loop signals and the burst occurrence times within compound ELMs.

Although the above is a conjecture, it frames hypotheses that are testable by direct time-domain analysis of the relevant signals if they can be obtained at sufficiently high time resolution, pointing
 to future work that may further the understanding of the ELMing process.

\section{Conclusions}

We have performed direct time domain analysis of ELMing in JET plasmas where a steady H-mode is sustained over several seconds, during which
 there is
no deliberate intent to control the ELMing process by external means. We identified the ELM occurrence times from the Be II signal, and have determined their relationship with
the phase of the VLD2 and VLD3 full flux loop  signals, which are a  high time resolution global measurement proportional to the voltage induced by changes
in poloidal magnetic flux in the divertor region.

 We have established that there is a signature of the build-up to an ELM in the phases of the full flux loop signals. Just before an ELM, the full flux loop phases progressively align such that at the ELM, they have the same value as at the previous ELM. This alignment is seen to develop over the $\sim 2-5 ms$ before the ELM. It is sufficiently strong  that it can be distinguished from phase relationships that could occur by coincidence in these quasi-oscillatory signals. We are able to recover this build-up signature using only data from before the ELM occurrence time. It thus possesses predictive power. While the full flux loops track each other at all times, that is, they have a phase relationship with each other that is distinct from random, they  become strongly phase synchronized within this build-up time before an ELM, consistent with globally spatially coherent plasma dynamics in the divertor region.

 These results may assist ELM prediction and mitigation, in that
 real time knowledge of the full flux loop signal phases
 indicates future times when ELM occurrence is statistically more likely.
The full flux loop signals capture aspects of the global dynamics of the plasma, including large scale plasma motion, plasma dynamics in the divertor region, and mutual interaction with the control system.
 Our result may thus provide new insight into the ELMing process. We suggest a possible scenario that unifies our understanding of intrinsic ELMing, and magnetic pacing of ELMs that uses the vertical stabilization coils to drive bulk plasma motion.

\begin{acknowledgments}
We acknowledge the EPSRC for financial support.
This work has been carried out within the framework of the EUROfusion Consortium and has received funding from the European Union's Horizon 2020 research and innovation programme under grant agreement number 633053, and from the RCUK Energy Programme [grant number EP/I501045]. The views and opinions expressed herein do not necessarily reflect those of the European Commission.
We acknowledge J. Davidsen and participants in the 2014 MPIPKS workshop on Causality, Information Transfer and Dynamical Networks for discussions, and A. J. Webster for provision of data.
\end{acknowledgments}

 \appendix*
 \section{Surrogate time series and null hypotheses}
The full flux loop signals can be seen in the time-series to have intervals where there is a clear sinusoidal component, with a characteristic period of $\sim 10ms$ and the ELM waiting times have time structure; they are not random. The analysis is performed on a restricted sized sample. We now test a series of null hypotheses that capture scenarios where the phase alignment that we report above could occur by coincidence. We will use the same circular statistics as above to distinguish the likelihood of coincidental occurrence in a quantitative manner. We use Hilbert transform window (a) and the same dataset as in the main paper, plasma 83771,  to construct the surrogates. We have repeated this analysis for all the other JET plasmas in this sequence, and we obtain the same results.

\subsection{ELM time and instantaneous phase of one full flux loop signal}

We need to quantify the phase alignment that could occur due to coincidence in comparing the ELM arrival times with a single signal, one of the full flux loops. If for example, the full flux loop signals were simply monochromatic sinusoids, and the ELMs occurred in a sufficiently periodic fashion, one would see ELMs preferentially occurring at particular phases in the full flux loop signals whether or not the sequence of ELM occurrence times and the full flux loop signals were related to each other.

\begin{figure}\centering
\includegraphics[scale=0.4]{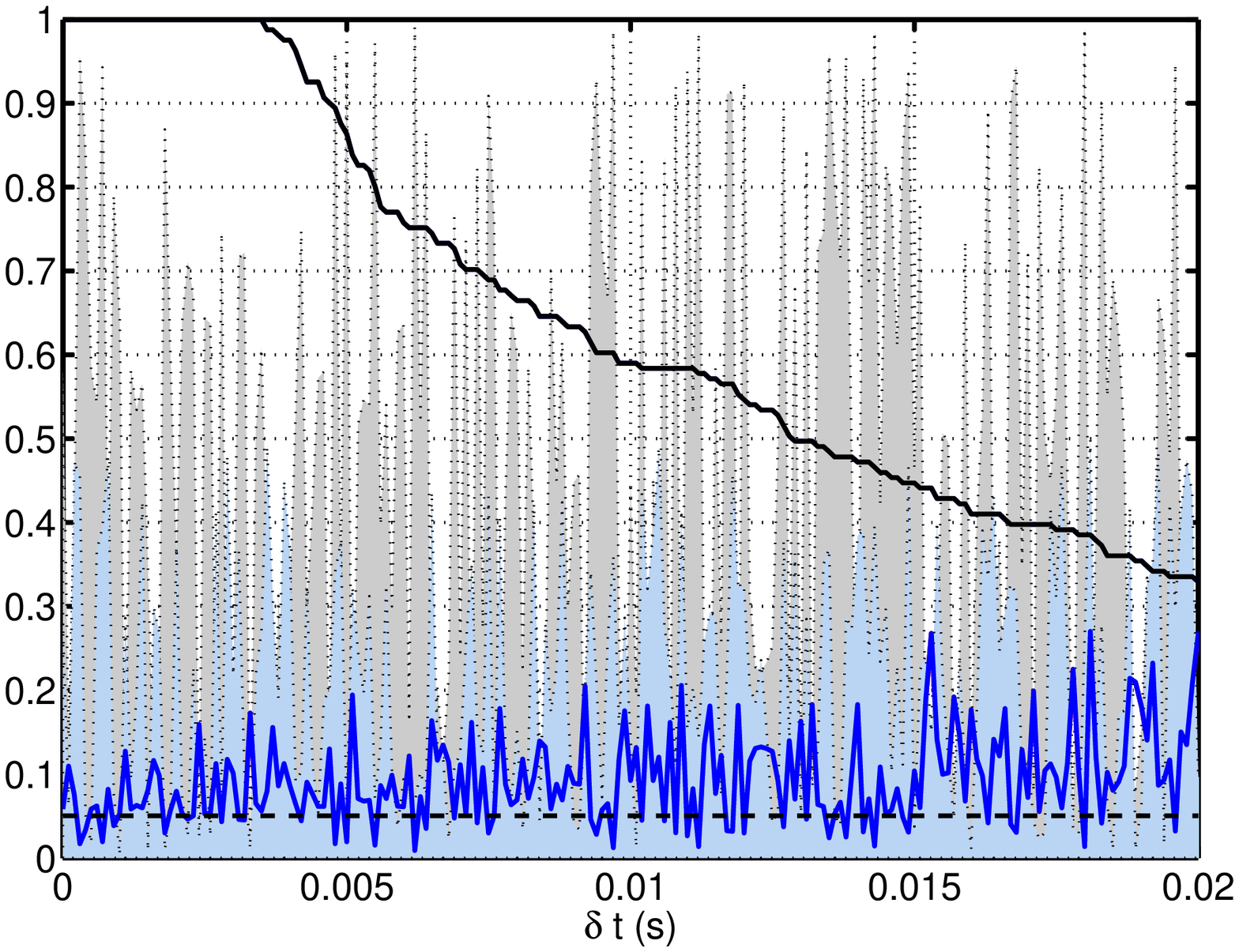}
\includegraphics[scale=0.4]{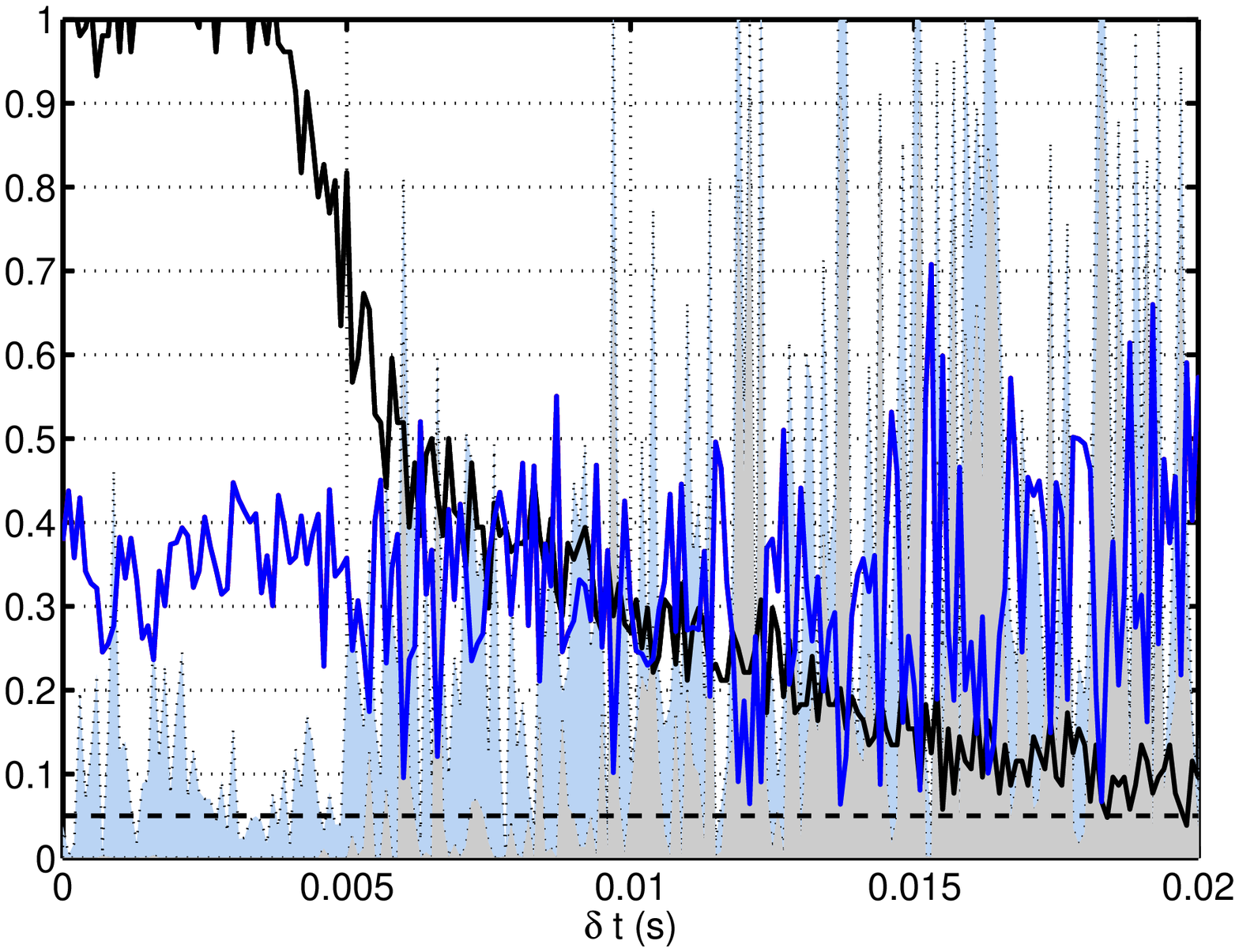}
\includegraphics[scale=0.4]{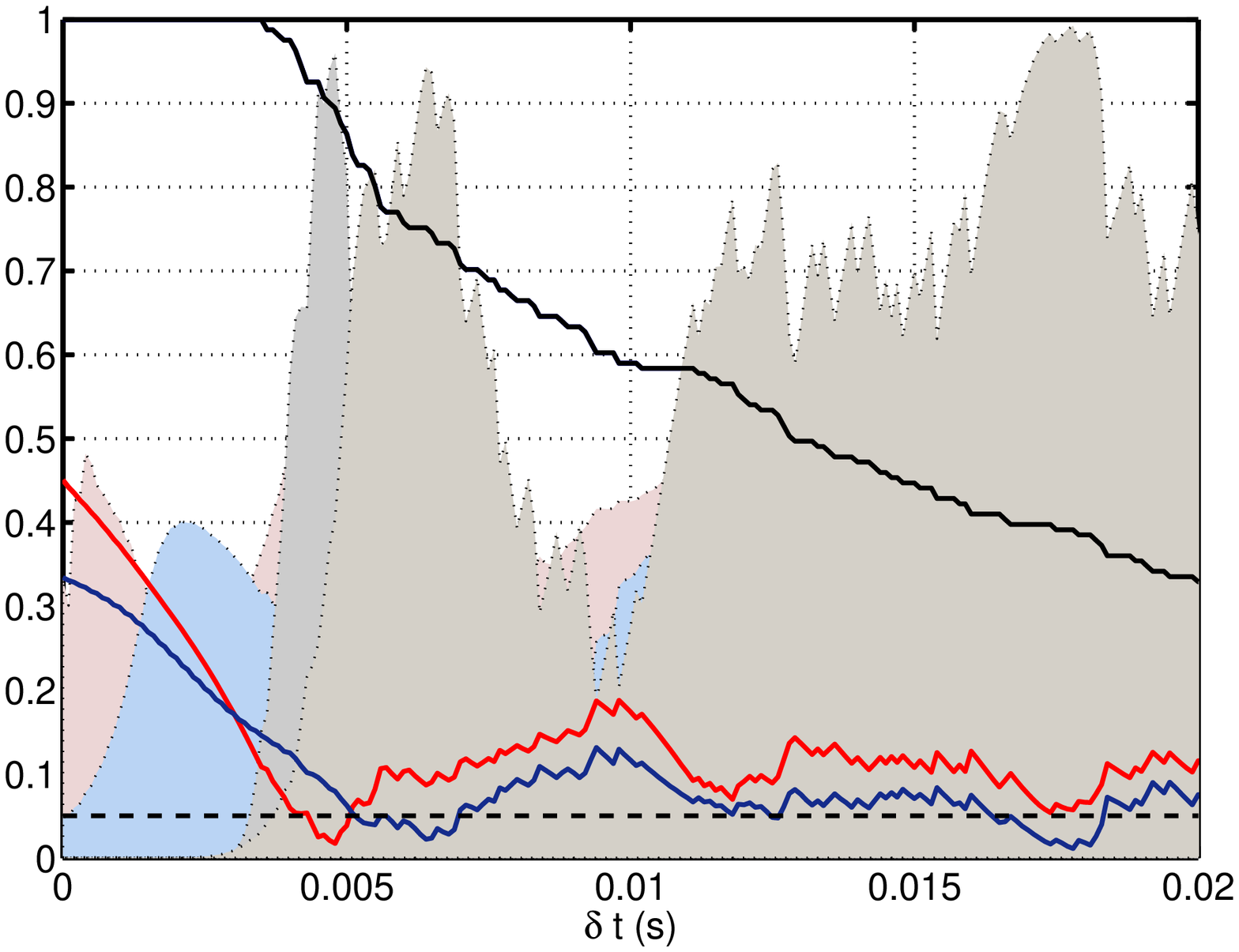}
\caption{Rayleigh statistics for VLD2 phase difference $\Delta \phi$ just before an ELM for three different surrogate time-series for the flat-top of JET plasma 83771. Format is as in previous figures. Top: the individual data points of the VLD2 signal have been randomly permuted within each ELM waiting time. Middle: the sequence of ELM waiting times is randomly permuted. Bottom: the VLD2 signal is replaced by a single sinusoid of period $T=10ms$ throughout the entire sequence of ELMs (blue); the sin phase is reset to zero at the time of the previous ELM (red). Red, blue shading the normalized angular variance. Light, dark grey the \emph{p}-value of single and reset sin waves respectively.
}\label{surrogateELMcoil}
\end{figure}
 We therefore test the statistical significance of the above results against some alternative hypotheses. We can represent these alternative hypotheses by constructing surrogate time-series that retain some, but not all, of the properties of the original data. We aim to test that the above results are significant compared to a random process. We also aim to quantify
 trivial correlation, that is, coincidences between ELM arrival time and full flux loop phase. Coincidences could arise in a finite data-set where both the sequence of
 ELM arrival times, and the full flux loop signals contain time structure that includes periodicity. Here, the ELM waiting times have a mean period and a 'comb like' multi-periodic structure, and the full flux loop signals exhibit intervals of oscillatory behaviour.
 We will calculate the same circular statistics in exactly the same manner as above for the following surrogate data-sets, the results are shown in the three panels of Figure \ref{surrogateELMcoil}.

 \textbf{\emph{A. No time correlation in the full flux loop data:}} Figure \ref{surrogateELMcoil} top panel. For each ELM pair we randomly permute (shuffle) the order of the full flux loop time series.

\textbf{\emph{B.  No pattern in the sequence of ELM waiting times:}} Figure \ref{surrogateELMcoil} middle panel. For each ELM waiting time $\Delta t_j$ we generate a surrogate ELM waiting time $\Delta t_s$ by selecting at random from  the time sequence of ELM waiting times $\{\Delta t_1, \Delta t_2,...\Delta t_j..\Delta t_N\}$, under the condition $\Delta t_s \leq \Delta t_j$.
 The surrogate set of ELM arrival times that this generates is
  $t_s=t_{j-1}+\Delta t_s$.  Each observed ELM pair then has a corresponding surrogate phase difference $\Delta \phi_s=\phi(t_s)-\phi(t_0)$ where $t_0$ is the arrival time of the first ELM and the second ELM has surrogate arrival time $t_s=t_0+\Delta t_s$.  The
  $\Delta t_s$ is drawn from the randomly permutated set of observed  ELM waiting times.

 \textbf{\emph{C. Full flux loops are single constant frequency sinusoids:}} Figure \ref{surrogateELMcoil} bottom panel. We replace the full flux loop signal with a sinusoid with period $T=10ms$, approximately the characteristic period of the oscillatory response seen following an ELM. We trial two surrogates, the first is a single sinusoid running through the entire time-series, and second, we reset the phase of the sinusoid to zero at the time if the first ELM in each pair, that is, at the start of each ELM waiting time.

From these surrogates we can conclude the following. First, Surrogate A establishes the value of $R \sim 0.1$ that occurs from the phases in a random signal. Here, $p>0.05$ so that the distribution of phases is indistinguishable from circular, they randomly occur at all angles. Surrogate B preserves both the full structure of the VLD2 signal and the probability distribution of ELM waiting times. Now, the ELM waiting time distribution has time structure, some waiting times occur more frequently than others. In a finite sized sample, randomly permuting them cannot generate coincidences with all possible phases of the VLD2 signal and this will lead to some alignment. As we move through the VLD2 time series by varying $\delta t$ the degree of alignment will fluctuate.
This can indeed be seen to give  $R\sim 0.3$  which is larger than the random signal surrogate A. In the $\delta t \sim 5ms$ before the surrogate ELM time, and both the angular variance $\sigma_\phi$ and  \emph{p}-value are small so that there is some alignment. This sets an upper bound for $R$ and a lower bound for $\sigma_\phi$ which can occur by such coincidences. Finally, surrogate C produces $p>0.05$ everywhere except $\delta t <3ms$. For $\delta t > 3ms$ the distribution of phases is indistinguishable from circular, they randomly occur at all angles. At smaller $\delta t$ there is again some alignment, which reaches a similar alignment, that is $R$ and angular variance $\sigma_\phi$, as in surrogate B, and for the same reason, the ELM waiting times have preferred values and these preferentially coincide with some phases of the single sinusoid surrogate.

Comparing these surrogates with our result of Figure \ref{elmallR}, we conclude that the alignment seen for $\delta t <5ms$ is statistically significant and cannot be accounted for by chance coincidence between the sequence of ELM occurrence times and the phase of the full flux loop signals. The alignment in  $5< \delta t <15ms$ is stronger than that of a random process ($R \sim 0.1$: surrogate A) but is comparable with  that arising from phase coincidence ($R\sim 0.3$, surrogates B and C) and thus cannot be distinguished from it.

\subsection{Phase difference between VLD2 and VLD3 signals}

The full flux loop signals both contain time structure that includes periodicity, on roughly the same period $T=10ms$. We now test against the coincidence that could occur in the phase difference between sinusoidal signals sampled at a sequence of times (the ELM arrival times) that have time structure.

\begin{figure}\centering
\includegraphics[scale=0.4]{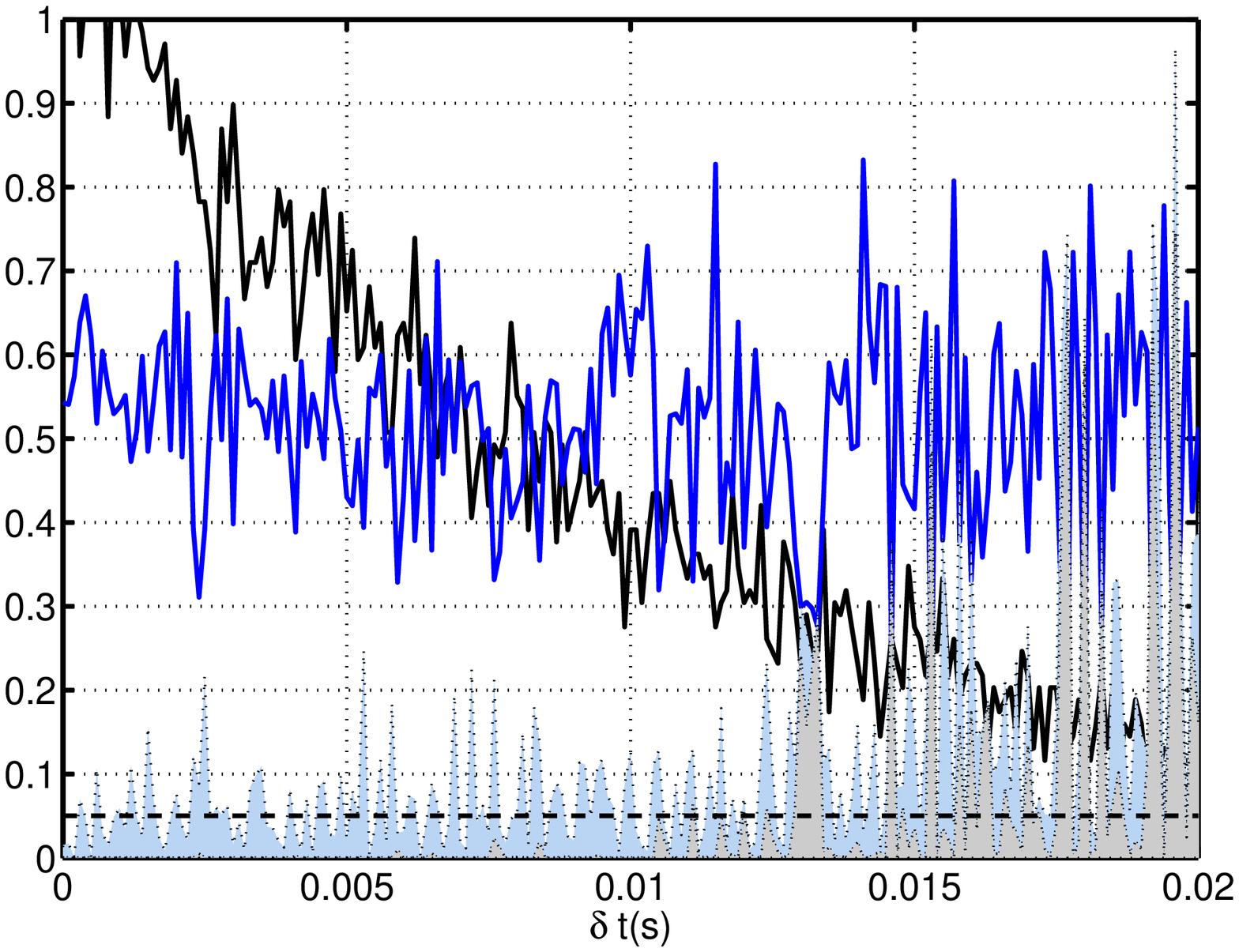}
\includegraphics[scale=0.4]{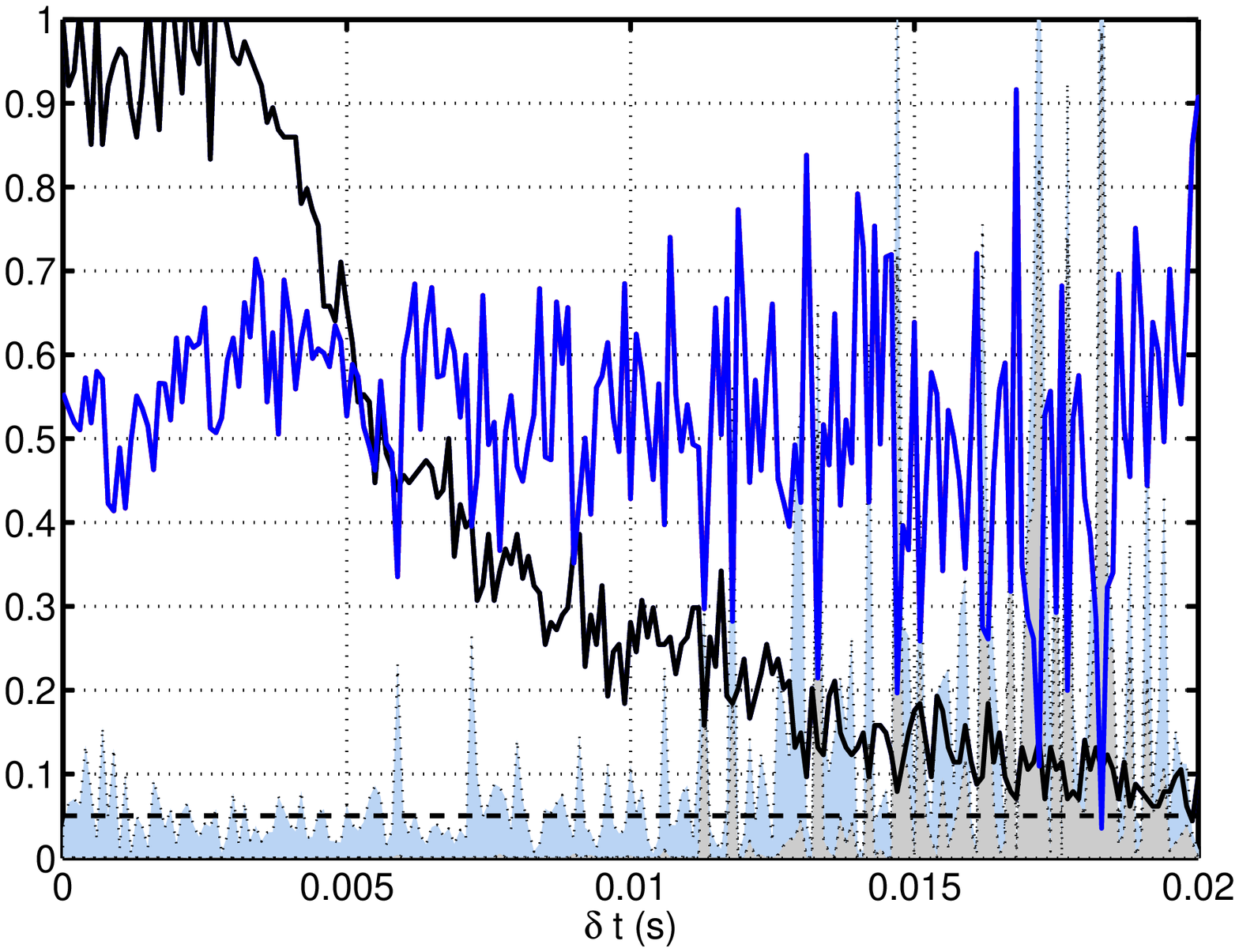}
\caption{Rayleigh statistics for phase difference between the VLD2 and VLD3 signals  just before an ELM for two different surrogate ELM time-series for the flat-top of JET plasma 83771. Format is as in previous figures. Here we use the unchanged VLD2 and 3 with surrogate EM arrival times. Top: the ELM arrives randomly at any time within the observed waiting time; Bottom: the ELM waiting times are randomly permuted.
}\label{surrogate2coilperm}
\end{figure}

\textbf{\emph{A. ELM arrives at a random time:}} Figure \ref{surrogate2coilperm} top panel. For each ELM pair we randomly select a time within the time interval to the next ELM, that is, the second ELM arrives at a random time.

\textbf{\emph{B.  No pattern in the sequence of ELM waiting times:}} Figure \ref{surrogate2coilperm} bottom panel. We randomly permute the time sequence of ELM waiting times  $\{\Delta t_1, \Delta t_2,...\Delta t_j..\Delta t_N\}$ as in surrogate B above.

\begin{figure}\centering
\includegraphics[scale=0.4]{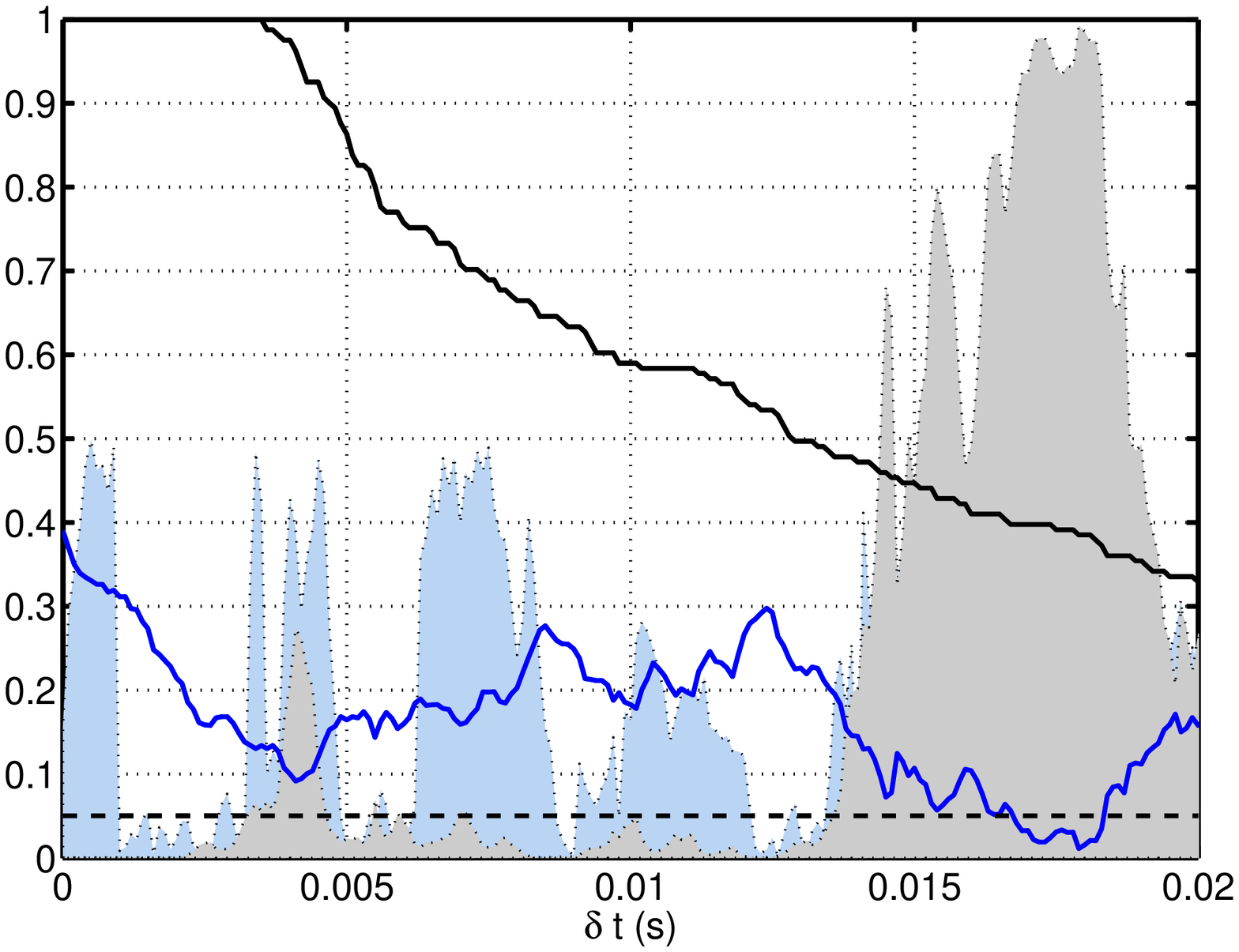}
\caption{Rayleigh statistics for phase difference between the VLD2 and VLD3 signals  just before an ELM for a surrogate VLD2 time-series for the flat-top of JET plasma 83771. Format is as in previous figures. Here we use the unchanged VLD3 and ELM arrival times. The VLD2 signal is replaced by a single sinusoid of period $T=10ms$ and the sin phase is reset to zero at the time of the previous ELM.
}\label{surrogate2coilsin}
\end{figure}

 \textbf{\emph{C. One of the full flux loops is a single constant frequency sinusoid:}} Figure \ref{surrogate2coilsin}. We replace the VLD2 full flux loop signal with a sinusoid with period $T=10ms$, the characteristic period of the oscillatory response seen following an ELM. We trial two surrogates, the first is a single sinusoid running through the entire time-series, and second, we reset the phase of the sinusoid to zero at the time if the first ELM in each pair, that is, at the start of each ELM waiting time. The results are similar, one case is shown.

These surrogates establish that the full flux loops are similar in phase at all times, surrogate A and B have an $R\sim 0.5$.  Comparing Figure \ref{clockdiffR} we see that this is the value at $\delta t>5ms$.


\begin{thebibliography}{20}

\bibitem{chappop}S. C. Chapman, R. O. Dendy, T. N. Todd, N. W. Watkins, A. J. Webster, F. A. Calderon et al., Phys. Plasmas 21 062302 (2014)
\bibitem{chap4pager}S. C. Chapman, R. O. Dendy, A. J. Webster, N. W. Watkins, T. N. Todd,
J. Morris and JET EFDA Contributors, 41st EPS Conference on Plasma Physics, June 2014,
Europhysics Conference Abstracts, 38F ISBN 2-914771-90-8, European Physical Society (2014)
http://ocs.ciemat.es/EPS2014PAP/pdf/P1.010.pdf

\bibitem{intro1}M. Keilhacker, Plasma Phys. Control. Fusion 26, 49
(1984).

\bibitem{intro2}V. Erckmann, F. Wagner, J. Baldzuhn, R. Brakel, R. Burhenn, U. Gasparino, P. Grigull, H. J. Hartfuss, J. V. Hofmann, R. Jaenicke et al.
Phys. Rev. Lett. 70, 2086 (1993).

\bibitem{intro3}H. Zohm, Plasma Phys. Control. Fusion 38, 105 (1996).

\bibitem{intro4}A. Loarte , G. Saibene, R. Sartori, D. Campbell, M. Becoulet, L. Horton, T. Eich, A. Herrmann, G. Matthews, N. Asakura, A. Chankin, A. Leonard, G. Porter, G. Federici, G. Janeschitz, M. Shimada, M. Sugiharaet, Plasma Phys. Control. Fusion 45, 1549
(2003).

\bibitem{intro5}K. Kamiya, N. Asakura, J. Boedo, T. Eich, G. Federici, M. Fenstermacher, K. Finken, A. Herrmann, J. Terry, A. Kirk, B. Koch, A. Loarte, R. Maingi, R. Maqueda, E. Nardon, N. Oyama. R. Sartori, Plasma Phys. Control. Fusion 49, S43
(2007).
\bibitem{Haw2009}R.J. Hawryluk, D.J. Campbell, G. Janeschitz, P.R. Thomas, R. Albanese, R. Ambrosino4, C. Bachmann, L. Baylor, M. Becoulet, I. Benfatto, J. Bialek  et al., Nucl. Fusion, 49, 065012, (2009)

\bibitem{peeling1}J. W. Connor,  Plasma Phys. Control Fusion 40 191 (1998)
\bibitem{peeling2} Snyder P.B., Wilson H.R., Ferron J.R., Lao L.L.,
Leonard A.W., Osborne T.H., Turnbull A.D.,
Mossessian D., Murakami M. and Xu X.Q., Phys.
Plasmas 9 2037 (2002)


\bibitem{kstar} G. S. Yun, W. Lee, M. J. Choi, J. Lee, H. K. Park, B. Tobias, C. W. Domier, N. C. Luhmann, Jr., A. J. H. Donné, J. H. Lee, and (KSTAR Team), Phys. Rev. Lett., 107, 045004 (2011)

\bibitem{deng}    A. W. Degeling, Y. R. Martin, P. E. Bak, J. B. Lister, and X. Llobet,
Plasma Phys. Cont. Fusion 43, 1671 (2001).


  \bibitem{greenh} J. Greenhough, S. C. Chapman, R. O. Dendy, and D. J. Ward, Plasma
Phys. Cont. Fusion 45, 747 (2003).
\bibitem{calderon} F. A. Calderon, R. O. Dendy, S. C. Chapman, A. J. Webster, B. Alper, R. M. Nicol and JET EDFA Contributors, Phys. Plasmas, 20 , 042306 (2013)

\bibitem{webster0}A. J. Webster and R. O .Dendy, Phys. Rev. Lett. 110, 155004 (2013)
\bibitem{webster}A. J. Webster, R. O.  Dendy, F.  Calderon, S. C. Chapman, E. Delabie, D. Dodt, R. Felton, T. Todd, V. Riccardo, B. Alper, S. Brezinsek,  et al, Plasma Phys. Control. Fusion, 56, 075017 (2014)
\bibitem{murari}A Murari, F Pisano, J Vega, B Cannas, A Fanni, S Gonzalez,
M Gelfusa, M Grosso and JET EFDA Contributors, Plasma Phys. Control. Fusion 56, 114007, (2014)

\bibitem{Mur} A. Murari, E. Peluso, P. Gaudio, M. Gelfusa, F. Maviglia, N. Hawkes and JET-EFDA Contributors, Plasma Phys. Control. Fusion 54, 105005 (2012)
\bibitem{gabor}D. Gabor,  Proc. IEE, vol. 93 (III),
pp. 429457, (1946)
\bibitem{text1} R. N. Bracewell, \emph{The Fourier transform and its applications, 2nd Ed.} (McGraw Hill, 1986)

\bibitem{phaseprl1} M. G. Rosenblum, A. S. Pikovsky, J. Kurths,  Phys. Rev. Lett, 76, 1804, (1996)




\bibitem{pikbook}A. Pikovsky, M. G. Rosenblum, J. Kurths,  \emph{Synchronization: a universal concept in nonlinear sciences}, C.U.P.,  (2003)

\bibitem{phaseprl2}J. T. C. Schwabedal, A. S. Pikovsky, Phys. Rev. Lett., 110, 24102, (2013)



\bibitem{fisher}N. I. Fisher,  \emph{Statistical Analysis of Circular Data}\emph{. Revised edition.} Cambridge University
Press.(1995)

\bibitem{berens}P. Berens, Journal of Stat. Software, 31,
September 2009, Volume 31, 10, (2009)
\bibitem{Luna2009} E.  de la Luna et al, Proceedings of the
36th EPS Conference on Plasma Physics, Sofia, Bulgaria.
29th June 2009 - 3rd July (2009)
\bibitem{sarkick}F. Sartori et al, 35th EPS Conference on Plasma Phys. Hersonissos, 9 - 13 June 2008 ECA Vol.32D, P-5.045 (2008)
 (2008)
\bibitem{langppcf}P. T. Lang et al, Plasma Phys. Cont. Fusion, 46, L31–L39, (2004)
\bibitem{nstx}S.P. Gerhardt, J-W. Ahn, J.M. Canik, R. Maingi, R. Bell,
D. Gates, R. Goldston, R. Hawryluk, B.P. Le Blanc,
J. Menard, A.C. Sontag, S. Sabbagh and K. Tritz, Nucl. Fusion, 50, 064015, (2010)
\bibitem{Neto2011} A. Neto et al, 2011 50th IEEE Conference on Decision and Control and
European Control Conference (CDC-ECC)
Orlando, FL, USA, December 12-15, (2011)

 \bibitem{kim09}S. H. Kim, M. M. Cavinato, V. Dokuka, A. A. Ivanov,
R. R. Khayrutdinov, P. T. Lang, J. B. Lister, V. E. Lukash, Y. R. Martin,
S. Yu Medvedev and L. Villard, Plasma Phys. Control. Fusion, 51, 055021 (2009)


\end{thebibliography}
\end{document}